\documentclass[english]{emulateapj}
\usepackage[T1]{fontenc}
\usepackage[latin1]{inputenc}
\usepackage{graphicx}
\usepackage{amssymb}

\makeatletter
\usepackage{graphicx}
\usepackage{amssymb}

\usepackage{graphicx}
\usepackage{amssymb}

\usepackage{times}
\usepackage{graphicx}
\usepackage{amssymb}
\usepackage{amsmath}

\newcommand{\yr}{{\,\rm yr}}
\newcommand{\pc}{\,\mathrm{pc}}

\newcommand{\mE}{\mathcal{E}}
\newcommand{\Mbh}{M_{\bullet}}

\newcommand{\Mo}{M_{\odot}}

\newcommand{\Rs}{R_{\star}}
\newcommand{\Ms}{M_{\star}}

\newcommand{\tp}{t_{\omega}}
\newcommand{\Jp}{J_{\omega}}

\bibpunct{\hspace{-2pt}}{}{}{a}{}{} 

\shorttitle{RESONANT RELAXATION}
\shortauthors{HOPMAN AND ALEXANDER}

\usepackage{babel}

\usepackage{babel}

\usepackage{babel}
\makeatother
\begin{document}

\title{Resonant relaxation near a massive black hole: \\
 the stellar distribution and gravitational wave sources}

\author{Clovis Hopman\altaffilmark{1} and Tal Alexander\altaffilmark{1,2}}

\altaffiltext{1}{Faculty of Physics, Weizmann
Institute of Science, POB 26, Rehovot 76100, Israel;
clovis.hopman@weizmann.ac.il, tal.alexander@weizmann.ac.il}
\altaffiltext{2}{Incumbent of the William Z. \& Eda Bess Novick career
development chair}



\email{clovis.hopman, tal.alexander@weizmann.ac.il}

\begin{abstract}
Resonant relaxation (RR) of orbital angular momenta occurs near
massive black holes (MBHs) where the potential is spherical and
stellar orbits are nearly Keplerian and so do not precess
significantly. The resulting \textit{coherent} torques efficiently
change the magnitude of the angular momenta and rotate the orbital
inclination in all directions. As a result, many of the tightly bound
stars very near the MBH are rapidly destroyed by falling into the MBH
on low-angular momentum orbits, while the orbits of the remaining
stars are efficiently randomized. We solve numerically the
Fokker-Planck equation in energy for the steady state distribution of
a single mass population with a RR sink term. We find that the steady
state current of stars, which sustains the accelerated drainage close
to the MBH, can be $\lesssim\!10$ larger than that due to
\textit{non-coherent} 2-body relaxation alone. RR mostly affects
tightly bound stars, and so it increases only moderately the total
tidal disruption rate, which is dominated by stars originating from
less bound orbits farther away.  We show that the event rate of
gravitational wave (GW) emission from inspiraling stars, originating
much closer to the MBH, is dominated by RR dynamics. The GW event rate
depends on the uncertain efficiency of RR. The efficiency indicated by
the few available simulations implies rates $\lesssim\!10$ times
higher than those predicted by 2-body relaxation, which would improve
the prospects of detecting such events by future GW detectors, such as
\textit{LISA.} However, a higher, but still plausible RR efficiency
can lead to the drainage of all tightly bound stars and strong
suppression of GW events from inspiraling stars. We apply our results
to the Galactic MBH, and show that the observed dynamical properties
of stars there are consistent with RR.
\end{abstract}


\section{Introduction}

Galactic nuclei with massive black holes (MBHs) are stellar systems
with relaxation times that are often shorter than the age of the Universe.
In that case the distribution function (DF) of the system may approach
a steady state. This steady state is determined by the boundary conditions
(an inner sink, such as the tidal radius of the MBH, and an outer
source at the interface with the host galaxy) and by the mutual interactions
between the stars themselves. The nature of the {}``microscopic''
interactions between the stars determines the rate at which the system
relaxes to its steady state.

With the notable exception of $N$-body simulations (e.g. Baumgardt,
Makino \& Ebisuzaki \cite{Baum04a}, \cite{Baum04b}; Preto, Merritt
\& Spurzem \cite{P04}; Merritt \& Szell \cite{MS05}), analyses of
the evolution of the DF near a MBH have almost exclusively relied
on the assumption that the mechanism through which stars exchange
angular momentum and energy is dominated by \textit{uncorrelated two-body
interactions} (Chandrasekhar \cite{Ch43}; see Binney \& Tremaine
\cite{BT87} for a more recent discussion). This assumption is made
in Fokker-Planck models (e.g. Bahcall \& Wolf \cite{BW76} {[}hereafter
BW76{]}; Bahcall \& Wolf \cite{BW77} {[}hereafter BW77{]}; Cohn \&
Kulsrud \cite{CK78}; Murphy, Cohn \& Durisen \cite{MCD91}), where
the microscopic interactions are expressed by the diffusion coefficients,
and in Monte Carlo simulations (e.g. Shapiro \& Marchant \cite{SM79};
Marchant \& Shapiro \cite{MS79}, \cite{MS80}; Freitag \& Benz \cite{FB01},
\cite{FB02}). Stars around MBHs are described as moving in the smooth
average potential of the MBH and the stars, and the scattering by
the fluctuating part of the potential is modeled as a hyperbolic Keplerian
interaction between a passing star and a test star. The scattering
effects accumulate non-coherently in a random-walk fashion.

The (non-resonant) relaxation time $T_{\mathrm{NR}}$ can be defined
as the time $T_{\mE}$ it takes for the negative specific energy $\mE\!\equiv\!-v^{2}/2-\varphi$
of a typical star (hereafter {}``energy'') to change by order unity.
This is also the time $T_{J}$ it takes for its specific angular momentum
$J$ (hereafter {}``angular momentum'') to change by an amount of
order $J_{c}(\mE)$, the maximal (circular orbit) angular momentum
for that energy%
\footnote{\label{ft:TJ}Throughout this paper, {}``angular momentum relaxation
time'' means the time it takes until $J$ is changed by order $J_{c}$,
rather than by order $J$. The time-scale for changes by order $J<J_{c}$
is shorter than the angular momentum relaxation time by a factor $(J/J_{c})^{2}$.%
}. In a spherical potential $J_{c}\!=\! G[\Mbh+N(>\!\mE)\Ms]/\sqrt{2\mE}$,
where $\Mbh$ is the MBH mass, $\Ms$ is the stellar mass and $N(>\!\mE)$
is the number of stars with orbital energies above $\mE$
(more bound than $\mE$). On Keplerian orbits $J_{c}\!=\!\sqrt{G\Mbh a}$,
where $a$ is the semi-major axis. When relaxation is dominated by
uncorrelated two-body interactions, the {}``non-resonant'' relaxation
time $T_{\textrm{NR}}$ of stars of mass $\Ms$ can be written in
the Keplerian regime as

\begin{equation}
T_{\textrm{NR}}=A_{\Lambda}\left({\frac{\Mbh}{\Ms}}\right)^{2}{\frac{P(a)}{N(<a)}}\qquad(\Mbh\!\gg\!\Ms)\,,\label{e:tr}\end{equation}
 where $P\!=\!2\pi\sqrt{a^{3}/(G\Mbh)}$ is the orbital period and
$A_{\Lambda}$ is a dimensionless constant which includes the Coulomb
logarithm. We assume throughout a single mass stellar population;
for numerical estimates we assume $\Ms\!=\!1\,\Mo$.

The assumption of uncorrelated two-body interactions is well-justified
in many systems, such as globular clusters. However, Rauch \& Tremaine
(\cite{RT96}, hereafter RT96) showed that this does not hold for
motion in potentials with certain symmetries, in particular for
Keplerian motion in the potential of a point mass where the orbits do
not precess because of the $1\!:\!1$ resonance between the radial and
azimuthal frequencies. This is to a good approximation the case for
stars orbiting close, but not too close to a MBH, since wider orbits
precess due to the potential of the enclosed stellar mass, while
tighter orbits precess due to General Relativity (GR). As long as the
precession timescale $\tp$ (the time to change the argument of the
periapse $\omega$ by $\pi$) is much longer than the orbital period,
the orbits effectively remain fixed in space over times $P\!\ll\!
t\!\ll\!\tp$ and the interactions are correlated. The orbiting stars
can then be represented by one-dimensional elliptical {}``wires''
whose mass density varies along the wire in proportion to the time
spent there (RT96). In this picture the interaction between two stars
can be described as the torque between two wires. This results in
mutual changes in \emph{both} the direction and the magnitude of the
angular momenta.  Following RT96, we denote such RR, which also
changes the magnitude of $\mathbf{J}$\emph{, scalar RR}. The change
grows coherently ($\propto\! t/\tp$) on timescales $t\!\ll\!\tp$ and
non-coherently ($\propto\!\sqrt{t/T_{\mathrm{RR}}}$) on longer
timescales $t\!\gg\!\tp$, where $T_{\mathrm{RR}}$ is the {}``resonant
relaxation'' (RR) timescale for the angular momentum
(Eq. \ref{e:TRR}). Generally, $T_{\mathrm{RR}}\ll\! T_{\mathrm{NR}}$
when $\Ms N(<\! a)\!\ll\!\Mbh$ (see Eq. \ref{e:TRRM}). Since the
potential of the wires is stationary on time-scales $\ll\!\tp$, they
do not exchange energy, and the energy relaxation timescale remains
long, $T_{\mE}\!\sim\! T_{\textrm{NR}}$, even in the resonant regime.
The mechanism of RR is reminiscent of the Kozai mechanism in triple
stars (Kozai \cite{KO62}).

A more restricted type of RR occurs in any spherical potential, where
the vector $\mathbf{J}$ is conserved, but $\omega$ precesses. In that
case the orbital rosette can be represented by a mass disk extending
between the orbital periapse and apo-apse. The mutual torques exerted
by such azimuthally symmetric disks change \emph{only} the direction
of $\mathbf{J}$, but not its magnitude. This type of RR persists even
in the presence of GR precession. Following RT96, we denote such RR,
which changes only the direction of $\mathbf{J}$, \emph{vector
RR}. The significance of the distinction between scalar and vector RR
is that scalar RR can deflect stars into {}``loss-cone'' orbits
($J\!\lesssim\! J_{c}\rightarrow J\!\sim\!0$) where they fall into the
MBH, either directly ({}``infall'') or gradually ({}``inspiral''),
whereas vector RR can only randomize the orbital orientation, but
cannot affect the loss rate.

In this paper we explore the consequences of RR on the stellar DF
and on the infall and inspiral rates. The mechanism of RR is briefly
reviewed in \S \ref{s:RR}. The Galactic Center (GC) model is defined
and the assumptions and approximations are discussed in \S \ref{s:model}.
In \S\ref{s:dens} We solve the Fokker-Planck equation in energy
and analyze the effects of RR on the DF (\S \ref{ss:DF}) and on
the stellar current (\S \ref{ss:flow}). In \S \ref{s:implications}
we discuss the observational consequences of RR, such as the event
rates of tidal disruption (\S\ref{sss:tiddisr}) and gravitational
wave (GW) emission (\S\ref{sss:GW}). We discuss our results in
the context of the young, non-relaxed stars in the GC (\S \ref{ss:GC}).
We discuss and summarize our results in \S \ref{s:summary}.

\section{Resonant relaxation}

\label{s:RR}

The resonant relaxation time $T_{\textrm{RR}}$ is estimated by
evaluating $\Delta\Jp$, the coherent change in the magnitude of the
specific angular momentum up to a time $\tp$. The change $\Delta\Jp$
is then the step size (``mean free path'') for the non-coherent growth
of the angular momentum over times $t\!>\!\tp$. Two nearby stars with
semi-major axes $a$ exert a mutual specific torque $\sim\!
G\Ms/a$. Within a distance $a$ from the MBH the net torque on a test
star fluctuates away from zero as $\dot{J}\!\sim\!\sqrt{N(<\!
a)}G\Ms/a$ and \begin{equation} \Delta\Jp\sim\dot{J}\tp=\sqrt{N(<\!
a)}(G\Ms/a)\tp\,.\label{e:Jw}\end{equation} For $t\!>\!\tp$ the
torques on a particular star-wire become random, and the change in
angular momentum grows in a random walk fashion with a timescale
$T_{\mathrm{RR}}\!\sim\!(J_{c}/\Delta J_{\omega})^{2}\tp$, defined
here as \begin{equation} T_{\textrm{RR}}\!\equiv\!
A_{\mathrm{RR}}\frac{N(>\!\mE)}{\mu^{2}(>\!\mE)}\frac{P^{2}(\mE)}{\tp}\!\simeq\!\frac{A_{\mathrm{RR}}}{N(<\!
a)}\left(\frac{\Mbh}{\Ms}\right)^{2}\frac{P^{2}(a)}{\tp}\,,\label{e:TRR}\end{equation}
where $\mu\!\equiv\! N\Ms/(\Mbh\!+\! N\Ms)$, $A_{\mathrm{RR}}$ is a
numerical factor of order unity, to be determined by simulations, and
the last approximate equality holds in the Keplerian regime. The
$N$-body simulations of RT96 indicate that
$A_{\mathrm{RR}}^{s}\!=\!3.56$ for scalar RR (\S\ref{ss:TRRs}) and
$A_{\mathrm{RR}}^{v}\!=\!0.31$ for vector RR (\S\ref{ss:TRRv}) (see
their Eqs. 8, 15 and table 4 for $\beta_{s}$ and $\beta_{v}$;
$A_{\mathrm{RR}}^{s,v}\!\equiv\!\beta_{s,v}^{-2}$).  We adopt these
values here. We verified these result, to within an order of unity, by
carrying a few full small scale $N$-body simulations.  Our simulations
exhibited a substantial scatter in the value of $A_{\mathrm{RR}}$, as
already noted by RT96. This introduces an uncertainty in quantitative
estimates of the efficiency of RR. Further detailed analysis and
simulations, outside the scope of this study, are needed to refine
these numerical estimates.

\subsection{Scalar resonant relaxation}

\label{ss:TRRs}

Over most of the relevant phase space, the precession is due to the
deviations from pure Keplerian motion caused by the potential of the
extended stellar cluster ({}``mass precession''). This occurs on
a timescale $\tp\!=\! t_{M}$, which assuming $N(<\! a)\Ms\!\ll\!\Mbh$
and after averaging on $J$, can be expressed as

\begin{equation}
t_{M}=A_{M}{\frac{\Mbh}{N(<\! a)\Ms}}P(a),\label{e:tprec}\end{equation}
 and where $A_{M}$ is a dimensionless constant. For simplicity we
adopt here $A_{M}\!=\!1$, the value for circular orbits. The $J$-averaged
RR timescale can then be written as

\begin{equation}
T_{\textrm{RR}}^{M}=A_{\mathrm{RR}}{\frac{\Mbh}{\Ms}}P(a)={\frac{A_{\mathrm{RR}}}{A_{\Lambda}}}{\frac{\Ms}{\Mbh}}N(<\! a)T_{\textrm{NR}}.\label{e:TRRM}\end{equation}
 Since $T_{\textrm{RR}}^{M}\!\ll\! T_{\textrm{NR}}$ for small $a$
where $N(<\! a)\Ms\!\ll\!\Mbh$, the RR rate of angular momentum relaxation
is much higher than the rate of energy relaxation in the resonant
regime. This qualitative analysis has been verified by detailed numerical
$N$-body simulations by RT96 and by Rauch \& Ingalls (\cite{RI98},
hereafter RI98).

For most of $J$-space, orbital precession is dominated by the mass
of the stellar cluster and the RR relaxation timescale is well approximated
by $T_{\textrm{RR}}\!\sim\! T_{\textrm{RR}}^{M}$, which is a function
of energy only. However, when the orbital periapse is very close to
the MBH, precession is dominated by GR effects ({}``GR precession'').
This is important for our analysis of the inspiral rate of GW sources
(\S\ref{sss:GW}). In this case the timescale for precession is given
by $\tp\!=\! t_{\mathrm{GR}}$, which is a strong function of $J$,

\begin{equation}
t_{\textrm{GR}}={\frac{8}{3}}\left(\frac{J}{J_{\textrm{LSO}}}\right)^{2}P,\label{e:tGR}\end{equation}
 where

\begin{equation}
J_{\textrm{LSO}}\equiv{\frac{4G\Mbh}{c}}\label{e:JLSO}\end{equation}
 is the angular momentum of the last stable orbit (LSO) for an orbit
of specific kinetic energy $\ll\! c^{2}$. Generally, both precession
mechanisms operate, and the scalar RR timescale $T_{\mathrm{RR}}^{s}(\mE,J)$
is given by substituting $1/\tp\!=\!\left|1/t_{M}-1/t_{\mathrm{GR}}\right|$
in Eq. (\ref{e:TRR}), where the opposite signs reflect the fact that
mass precession is retrograde whereas GR precession is prograde. Thus,
the scalar RR timescale is\begin{equation}
T_{\mathrm{RR}}^{s}=\frac{A_{\mathrm{RR}}}{N(<\! a)}\left(\frac{\Mbh}{\Ms}\right)^{2}P^{2}(a)\left|\frac{1}{t_{M}}-\frac{1}{t_{\mathrm{GR}}}\right|\,,\label{e:TRRs}\end{equation}

When $t_{\mathrm{GR}}\!\ll\! t_{M}$ and GR precession dominates,
the RR timescale is (Eq. \ref{e:TRRM}) \begin{equation}
T_{\textrm{RR}}^{\textrm{GR}}=\frac{3}{8}A_{\mathrm{RR}}\left(\frac{\Mbh}{\Ms}\right)^{2}\left(\frac{J_{\mathrm{LSO}}}{J}\right)^{2}\frac{P(a)}{N(<\! a)}\,.\label{e:TRRGR}\end{equation}

The fact that scalar RR becomes much less efficient due to GR precession
is crucial for the viability of GW emission from inspiraling stellar
objects ({}``extreme mass-ratio inspiral sources'', EMRIs), since
it allows compact remnants to be very rapidly deflected to strongly
relativistic orbits, but then stall {}``on the brink'' and instead
of falling directly into the MBH, inspiral into it gradually by the
emission of GW. We return to this issue in \S\ref{sss:GW}.

We analyze below the dynamics of scalar RR and its effects on the
stellar DF, both numerically (\S\ref{s:dens}) and by Monte-Carlo
simulations (\S\ref{s:implications}). These two approaches require
different formulations of the RR timescale. The numerical analysis
is carried out in the 1D $\mE$-space, where the $J$-dependences
are absorbed in the $J$-averaged terms. For this purpose we use the
relation $\mathrm{d}(J^{2})/J_{c}^{2}\!=\!\mathrm{d}t/T_{\mathrm{RR}}^{s}(\mE,J)$
(Eqs. \ref{e:TRRs}) to define the $J$-averaged time it takes a star
to random-walk from $J=J_{c}(\mE)$ to the loss-cone $J=J_{lc}$ as

\begin{equation}
\bar{T}_{\textrm{RR}}^{s}(\mE)=\frac{1}{J_{c}^{2}}\int_{J_{lc}^{2}}^{J_{c}^{2}}dJ^{2}T_{\textrm{RR}}^{s}(\mE,J)\,.\label{e:Tave}\end{equation}
 Since at low energies $\bar{T}_{\textrm{RR}}^{s}\!\propto P$, it
initially decreases with $\mE$ toward the MBH. However, at very large
energies near the MBH, where GR precession becomes important ($J_{\mathrm{LSO}}/J_{c}(\mE)\!\rightarrow\!1$),
$\bar{T}_{\textrm{RR}}^{s}$ increases again (Fig. \ref{f:GC}). The
Monte Carlo simulations are carried out in 2D ($\mE$,$J$)-space.
For this purpose we define a general expression for the scalar angular
momentum relaxation time, $T_{J}^{s}$, which applies in any regime
(Eqs. \ref{e:tr}, \ref{e:TRR}, \ref{e:TRRs}; see also footnote \ref{ft:TJ}).

\begin{equation}
T_{J}^{s}(\mE,J)=\left[\frac{1}{T_{\textrm{NR}}(\mE)}+\frac{1}{T_{\textrm{RR}}(\mE,J)}\right]^{-1}\,.\label{e:TJ}\end{equation}

\subsection{Vector resonant relaxation}

\label{ss:TRRv} 

Vector RR grows coherently ($\propto\! t$) on timescales $t\!\ll\!
t_{\varphi}$, where $t_{\varphi}$ is the timescale for a change of
order unity in the total gravitational potential $\varphi$ caused by
the changes in the stellar potential $\varphi_{\star}$ due to the
realignment of the stars as they rotate by $\pi$ on their
orbits,\begin{equation}
t_{\varphi}=\frac{\varphi}{\dot{\varphi_{\star}}}=A_{\varphi}\frac{N^{1/2}}{\mu}\frac{P}{2}\simeq\frac{1}{2}A_{\varphi}\frac{\Mbh}{\Ms}\frac{P}{N^{1/2}}\,,\label{e:tphi}\end{equation}
where $A_{\varphi}$ is a dimensionless constant of order unity, and
where the last approximate equality holds for $N\Ms\!\ll\!\Mbh$.  For
simplicity we adopt here $A_{\varphi}\!=\!1$. In analogy to scalar RR
(Eq. \ref{e:Jw}), the maximal coherent change in $\mathbf{J}$ is
\begin{equation}
\left|\Delta\mathbf{J}_{\varphi}\right|\sim\dot{J}t_{\varphi}\sim
J_{c}\,,\end{equation} that is, $\mathbf{J}$ rotates by an angle
${\cal {O}}(1)$ already during the coherent phase. On timescales
$t\!\gg\! t_{\varphi}$, $\left|\Delta\mathbf{J}_{\varphi}\right|$
cannot grow larger, as it already reached its maximal possible value,
but the orbital inclination angle is continuously randomized
non-coherently ($\propto\! t^{1/2}$) on the vector RR timescale
(Eq. \ref{e:TRR}),

\begin{equation}
T_{\mathrm{RR}}^{v}=2A_{\mathrm{RR}}^{v}\frac{N^{1/2}(>\!\mE)}{\mu(\mE)}P(\mE)\simeq2A_{\mathrm{RR}}^{v}\left(\frac{\Mbh}{\Ms}\right)\frac{P(a)}{N^{1/2}(<\! a)},\label{e:Tv}\end{equation}
 where the last approximate equality holds for $N\Ms\!\ll\!\Mbh$. 

Note that while the torques driving scalar and vector RR are the
same, vector RR is much more efficient than scalar RR, $T_{\mathrm{RR}}^{v}\!\ll\!\bar{T}_{\mathrm{RR}}^{s}$,
due to the much longer coherence time $t_{\varphi}\!\sim\! N^{1/2}t_{M}\!\gg\! t_{M}$.
Furthermore, vector RR proceeds irrespective of any precession mechanisms
that limit the efficiency of scalar RR.

\section{A stellar cluster with a massive black hole: model and assumptions}

\label{s:model}

In this section we state our assumptions, approximations and definitions,
many of which are commonly used in the analysis of stellar systems
with MBHs. Since the GC is the best studied case of a stellar cluster
with a MBH, and since it is representative of typical \textit{LISA}
targets, we scale the parameters to values appropriate for this system.
Our goal is to find the stellar DF in presence of RR. A full solution
in ($\mE,J$)-space is complex and is not attempted here. For simplicity,
the analysis carried out here is in $\mE$-space only and it assumes
an underlying spherical symmetry. The $J$-dependent loss-cone terms are
approximated by their $J$-averaged effective terms, which are functions
of $\mE$ only.

\subsection{Distribution in energy space}

\label{ss:energyDF}

Recent observations show that a tight empirical relation exists between
the mass of MBHs, $\Mbh$, and the velocity dispersion $\sigma$ of
their host bulge or galaxy (Ferrarese \& Merritt \cite{FM00}; Gebhardt
et al. \cite{Geb00}),

\begin{equation}
M_{\bullet}=1.3\times10^{8}M_{\odot}\left(\frac{\sigma}{200\,\mathrm{km\, s^{-1}}}\right)^{4}\,,\label{e:Msigma}\end{equation}
 where $\sigma$ is measured at the galaxy's effective radius (Tremaine
et al. \cite{Tr02}). Both the exponent and the pre-factor have small
uncertainties, which we ignore here. The Galaxy obeys the $\Mbh$-$\sigma$
relation, with $\Mbh\!\sim\!3\!\times\!10^{6}\Mo$, and $\sigma\!\sim\!75\,\mathrm{km\, s^{-1}}$
(Tremaine et al. \cite{Tr02}).

The MBH dominates the dynamics of stars within its {}``Bondi radius'',
or radius of influence,

\begin{eqnarray}
r_{h} & = & {\frac{G\Mbh}{\sigma^{2}}}=2\,\mathrm{pc}\,\left({\frac{\Mbh}{3\times10^{6}\Mo}}\right)\left({\frac{\sigma}{75\,\mathrm{km\, s^{-1}}}}\right)^{-2}\label{e:rh}\\
 & = & 2\,\mathrm{pc}\,\left({\frac{\Mbh}{3\times10^{6}\Mo}}\right)^{1/2},\end{eqnarray}
 where in the last step we used the $\Mbh$-$\sigma$ relation. Within
$r_{h}$ the system is approximately Keplerian, and deviations from
Keplerian motion become important only for times $t\!\gg\! P$, where
$P(\mE)=2\pi\sqrt{(G\Mbh)^{2}/8\mE^{3}}$ is the orbital period and
the energy is $\mE=G\Mbh/r-v^{2}/2$.

We assume a single mass population with an underlying spherically
symmetric DF, so that $f(\mathbf{x},\mathbf{v})\!=\! f(\mE)$ (stars
per phase space volume $\mathrm{d}^{3}x\mathrm{d}^{3}v$). Following
BW76 we assume that for $r\!>\! r_{h}$ the DF is Maxwellian,

\begin{equation}
f(\mE)={\frac{n_{h}}{(2\pi\sigma_{h}^{2})^{3/2}}}e^{\mE/\sigma_{h}^{2}}\qquad(r>r_{h})\,,\label{e:MB}\end{equation}
where $n_{h}\approx4\times10^{4}\,\mathrm{pc^{-3}}$ is the number
density at $r_{h}$ in the GC assuming a mean mass of $1\,\Mo$ (Genzel
et al. \cite{Gea03}) and where $\sigma_{h}\!=\!75\,\mathrm{km\, s^{-1}}$
is the 1D velocity dispersion at $r_{h}$. For Keplerian orbits, the
density $n(\mE)$ (stars per $\mathrm{d}\mE$) is

\begin{equation}
n(\mE)=\pi^{3}\sqrt{2}(G\Mbh)^{3}\mE^{-5/2}f(\mE)\,,\label{e:NE}\end{equation}
 the density $n(r)$ (stars per $\mathrm{d^{3}}x$) is

\begin{equation}
n(r)\!=\!4\sqrt{2}\pi\int_{-\infty}^{G\Mbh/r}\!\mathrm{d}\mE f(\mE)\sqrt{{\frac{G\Mbh}{r}}-\mE}\,,\label{e:nr}\end{equation}
 and the density $n(a)$ (stars per $\mathrm{d}a$) is

\begin{equation}
n(a)=4\pi^{3}(G\Mbh)^{3/2}a^{1/2}f\left({\frac{G\Mbh}{2a}}\right)\,.\label{e:na}\end{equation}

Two stellar components contribute to the local stellar density at
radius $r$. One is the tightly bound stars with energy $\mE\!\sim\!{\mathcal{O}}(G\Mbh/r)$,
or $a\!\sim\!{\mathcal{O}}(r)$. The other is the unbound, or marginally
bound stars with energy $\mE\!\ll\! G\Mbh/r$ (or $a\!\gg\! r$) and
high eccentricities, which spend only a small fraction of their orbit
inside $r$. When unbound stars dominate the local population, for
example because RR or stellar collisions have destroyed most of the
tightly bound stars there, the density profile is given by

\begin{equation}
n_{u}(r)\!=\!4\sqrt{2}\pi\int_{-\infty}^{0}\!\mathrm{d}\mE f(\mE)\sqrt{{\frac{G\Mbh}{r}}-\mE}\propto r^{-1/2}.\label{e:nru}\end{equation}

We express the local relaxation time, which generally depends on the
nature of the relaxation process and on radius, in terms of a reference
time $T_{h}$, which we define as the NR relaxation time at $r_{h}$,

\begin{eqnarray}
T_{h} & \equiv & {\frac{3}{16}}\sqrt{{\frac{2}{\pi}}}{\frac{\sigma_{h}^{3}}{n_{h}(G\Ms)^{2}\ln\Lambda}}\nonumber \\
 & = & 6\times10^{9}\yr\left({\frac{\Mbh}{3\times10^{6}\Mo}}\right)^{3/4},\end{eqnarray}
 where the last equality assumed the $\Mbh$-$\sigma$ relation and
the stellar density in the GC. The Coulomb factor is $\Lambda\!=\! r_{\max}/r_{\min}\!=\!\Mbh/\Ms$
(BW76), where $r_{\max}\!\sim\! r$ is the maximal impact parameter
for perturbations by stars interior to $r$, and $r_{\min}$ is the
minimal impact parameter for small deflections, $r_{\min}=G\Ms/v^{2}\!=\!(\Ms/\Mbh)r$.
The local NR relaxation time may be quite different from $T_{h}$
when the density distribution of the system strongly deviates from
the steady state configuration $n(r)\propto r^{-7/4}$ (\S\ref{ss:DF}).

\subsection{The loss-cone }

\label{ss:losscone}

At high enough energy the DF vanishes, $f(\mE\!>\!\mE_{D})\!=\!0$,
because stellar objects cannot survive close to the MBH. The value of
$\mE_D$ depends on the process that destroys the stars. If they are
compact objects, they fall directly into the MBH or inspiral into it
by GW emission. If they are main sequence stars, they are destroyed by
disruptive stellar collisions in the high density cusp around the MBH,
or are tidally captured and heated until disruption, or are tidally
disrupted when their orbital periapse $r_{p}$ falls below the tidal
disruption radius, $r_{t}\simeq\left(\Mbh/\Ms\right)^{1/3}\Rs$.

Stars are destroyed by the MBH either by scattering or decaying in
$\mE$-space to the point where $\mE\!>\!\mE_{D}$, or by scattering in
$J$-space to the point where $J\!<\! J_{lc}\equiv\sqrt{2G\Mbh r_t}$
(or $r_{p}\!<\! r_{t}$).  Scattering in $J$-space is by far more
efficient (Frank \& Rees \cite{FR76}; Lightman and Shapiro
\cite{LS77}; see footnote \ref{ft:TJ}), and the tidal disruption rate
is dominated by stars on low-$\mE$ and low-$J$ orbits. Stars enter the
loss-cone, defined by $J<J_{lc}$, in two regimes of phase-space,
depending on their orbital energy, which determines the ratio between
the angular momentum change per orbit, $\Delta J(\mE)$, and $J_{lc}$.
For small energies (long periods), $\Delta J\!\gg\! J_{lc}$. In this
{}``full loss-cone'' (or {}``kick'') regime, the DF remains
essentially unmodified by the existence of a loss-cone. For high
energies (short periods) near the MBH, $\Delta J\!\ll\! J_{lc}$. In
this {}``empty loss-cone'' (or {}``diffusive'') regime stars slowly
diffuse into the loss-cone and are then promptly destroyed on a
dynamical time, and the DF is modified by the existence of the
loss-cone. The stellar density vanishes inside the loss-cone,
$n(\mE,J\!<\! J_{lc})\!=\!0$, and gradually falls to zero toward it,
$n(J)\propto J\ln(J/J_{lc})$ for $J\gtrsim J_{lc}$ (Lightman \&
Shapiro \cite{LS77}). This also modifies somewhat the DF in
$\mE$-space. Note that scalar RR is typically efficient only deep in
the empty loss-cone regime, where mass precession is negligible.

There is a critical energy where $\Delta J(\mE)\!=\! J_{lc}$, that
demarcates the transition between the empty and full loss-cone regimes.
The size of the loss-cone, and therefore the critical energy, depends
on the nature of the loss-process. The critical energy for prompt
stellar disruption $\mE_{p}$, corresponds to a distance scale of
$\sim\! r_{h}$ where $\mE\!=\!\mE_{h}$ (Eq. \ref{e:rh}). Most of
the contribution to the stellar destruction rate is from stars with
$\mE\!\gtrsim\!\mE_{p}$. Similarly, there is a critical energy for
stellar destruction by GW inspiral, $\mE_{\textrm{GW}}$. GW inspiral
takes much longer than direct disruption. Unless the star starts out
on a short-period (high-$\mE$) orbit, it will be scattered again
directly into the MBH, or to a high-$J$ orbit where GW dissipation
is negligible. Thus, $\mE_{\mathrm{GW}}\!\ll\!\mE_{p}$ (Alexander
\& Hopman \cite{AH03}). 

We adopt below for numerical calculations the values of the critical
energies in the GC, where for consistency with our assumed single
mass population, we consider only prompt disruption and GW inspiral
at the last stable orbit, $J_{lc}\!=\!4G\Mbh/c$. In the GC, the semi-major
axes corresponding to these critical energies are $a_{p}\!=\!0.27\,\mathrm{pc}$
and $a_{GW}\!=\!0.02\,\mathrm{pc}$ (see Fig. \ref{f:Sa}). Thus,
our model describes a population of compact objects, and our predicted
prompt disruption rates will be slightly lower than the actual tidal
disruption rates, which occur at the somewhat larger $J_{lc}\!=\!\sqrt{2G\Mbh r_{t}}$
and thus have a lower critical energy, closer to $\mE_{h}$.

A complete treatment of the loss-cone problem involves the solution
of the Fokker-Planck equation in $(\mE,J)$-space (Cohn \& Kulsrud
\cite{CK78}). However, an approximate solution can be obtained by
solving the Fokker-Planck equation in $\mE$-space only, while accounting
for the loss-cone by adding sink terms to the equation (BW77). This
is the approach we adopt here. The tidal disruption rate due to NR
relaxation has been discussed extensively in the literature (see e.g.
Lightman and Shapiro \cite{LS77}; Frank \& Rees \cite{FR76}; Cohn
\& Kulsrud \cite{CK78}; Syer \& Ulmer \cite{SU99}; Magorrian \&
Tremaine \cite{MT99}; Alexander \& Hopman \cite{AH03}; Merritt \&
Poon \cite{MP04}; Wang \& Merritt \cite{WM04}; Hopman \& Alexander
\cite{HA05}; Baumgardt et al. \cite{Bea05}). The differential NR
loss-cone diffusion rate (stars per $\mathrm{d}\mE\mathrm{d}t$) is
estimated to be of the order

\begin{equation}
\Gamma_{\textrm{NR}}(\mE)\sim n(\mE)/T_{\textrm{NR}}(\mE)\,,\label{e:NRlc}\end{equation}
 neglecting here the weak logarithmic $J$-dependence that expresses
the depletion of phase space near the loss-cone. When RR is the dominant
relaxation process, then for timescales $\!\gg\!\tp$ the differential
loss-cone refilling rate is%
\footnote{Eq. (\ref{e:RRlc}) differs from the corresponding term in RT96, which
applies only for timescales $\ll\!\tp$, and an isotropic DF. The rate
predicted by Eq\@. (\ref{e:RRlc}) is in good agreement with the
values obtained by the $N$-body simulations of RI98. %
}

\begin{equation}
\Gamma_{\textrm{RR}}(\mE)\sim n(\mE)/\bar{T}_{\textrm{RR}}^{s}(\mE)\,.\label{e:RRlc}\end{equation}
Note that vector RR does not enter into the Fokker Planck equation
and does not play a role in the diffusion in $\mE$-space since the
system is assumed to be spherically symmetric, and vector RR does
not change $J$.

\section{Density profile near a massive black hole}

\label{s:dens}

\subsection{Fokker-Planck energy equation with resonant relaxation}

\label{ss:FP}

Following BW76, we define the dimensionless quantities

\begin{equation}
g=(2\pi\sigma_{h}^{2})^{3/2}n_{h}^{-1}f;\qquad x=\mE/\sigma_{h}^{2};\qquad\tau=t/T_{h}\,,\label{e:g}\end{equation}
 and write the Fokker-Planck equation%
\footnote{Equation (\ref{e:dgdt}) is written in the form of a particle conservation
relation, which, as shown by BW76, is equivalent to the usual Fokker-Planck
form $\partial f/\partial t=-D_{x}\partial f/\partial x+D_{xx}\partial^{2}f/\partial x^{2}$,
where $D_{x}$ and $D_{xx}$ are the diffusion coefficients. %
} as

\begin{equation}
{\frac{\partial g(x,\tau)}{\partial\tau}}=-x^{5/2}{\frac{\partial Q(x,\tau)}{\partial x}}-R_{\textrm{NR}}(x,\tau)-\chi R_{\mathrm{RR}}(x,\tau),\label{e:dgdt}\end{equation}
 where the current $Q(x)$, defined to be the net rate at which stars
flow to energies $>x$, is given by (BW76)\begin{eqnarray}
Q(x,\tau)=\int_{-\infty}^{x_{D}} & dy & \left[g(x,\tau){\frac{\partial g(y,\tau)}{\partial y}}-g(y,\tau){\frac{\partial g(x,\tau)}{\partial x}}\right]\times\nonumber \\
 & \times & \left\{ \max(x,y)\right\} ^{-3/2}.\end{eqnarray}
The sink terms $R_{\mathrm{NR}}$ and $R_{\mathrm{RR}}$ represent
the $J$-averaged differential loss rate of stars per energy interval,
by NR and RR respectively, due the existence of a loss-cone in $J$-space.
This is an approximate substitute for the full $2\!+\!1$ treatment.

The NR sink term $R_{\textrm{NR}}(x,\tau)$ is expressed by the form
(Eq. \ref{e:NRBW} in appendix B) proposed by BW77, which smoothly
interpolates between the empty and the full loss-cone regimes and
approximates the logarithmic depletion of phase space near the loss-cone
boundary (\S\ref{ss:losscone}). In the empty loss-cone regime, which
is of interest here, \begin{equation}
R_{\textrm{NR}}(x,\tau)\!\propto\! g(x,\tau)^{2}\left/\log(x_{D}/4x)\right.\,.\label{e:RNR}\end{equation}
 This quadratic behavior arises since a fraction $g(x,\tau)$ of the
stars is accreted per local relaxation time, which is itself proportional
to $1/g(x,\tau)$.

The RR sink term $R_{\mathrm{RR}}$ is expressed by (Eqs. \ref{e:TRRM},
\ref{e:RRlc}) \begin{equation}
R_{\mathrm{RR}}(x,\tau)=g(x,\tau)\left/\tau_{\textrm{RR}}(x)\right.\,,\label{e:RRR}\end{equation}
 where $\tau_{\textrm{RR}}(x)\!\equiv\!\bar{T}_{\mathrm{RR}}^{s}(\mE)/T_{h}$
is the dimensionless $J$-averaged RR time, defined in Eq. (\ref{e:Tave}).

The RR efficiency factor $\chi$ in Eq. (\ref{e:dgdt}) parametrizes the
uncertainties in the efficiency of RR and in the approximations
involved in Eq. (\ref{e:RRR}), for example the effect of the partial
depletion of phase-space outside the loss-cone (note that the
depletion is small, see RI98 fig. [2a]). For simplicity, we assume
that $\chi$ is a constant. We expect ${\mathcal{\chi\!\sim\!  O}}(1)$,
and show below that the choice $\chi\!=\!1$ yields results that are
consistent with those of RT96 and RI98, but we also explore RR for a
range of $\chi$ values to determine the robustness of our results.
The RR sink term becomes negligible compared to the NR sink term for
lower energies, were $T_{\textrm{RR}}\!\gg\! T_{h}$ (Fig. \ref{f:GC}).

We follow BW76 by assuming as boundary conditions

\begin{equation}
g(x\!<\!0,\tau)=e^{x}\,,\qquad g(x\!=\! x_{D},\tau)=0.\label{e:BC}\end{equation}
 The first boundary condition for $x\!<\!0$ expresses the assumption
of a thermal reservoir at the radius where the stars are no longer
bound to the MBH (Eq. \ref{e:MB}). In reality, they are bound to
the total enclosed mass of the MBH and stars, however the BW76 treatment
neglects the stellar mass and assumes Keplerian motion around the
MBH. The second expresses the existence of a mass-sink at energy $x_{D}$.
Because of the large dynamical range in the problem, it is more convenient
for numerical purposes to express Eqs. (\ref{e:dgdt}-\ref{e:BC})
in logarithmic energy (see appendix \ref{s:LogE}).

\subsection{The effects of RR on the stellar distribution function}

\label{ss:DF}

\begin{figure}
\includegraphics[%
  width=1.0\columnwidth,
  keepaspectratio]{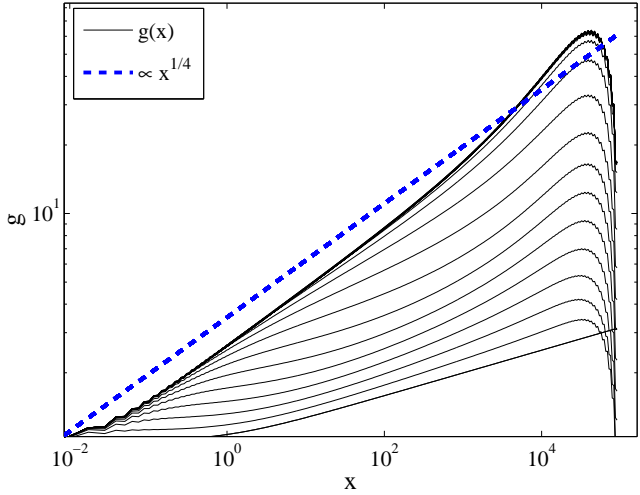}

\caption{\label{f:bw76} The evolution of the stellar DF to a steady-state
$\mE^{1/4}$ cusp (BW76) by $\mE$-diffusion only (no loss-cone effects:
$R_{\textrm{NR}}\!=\! R_{\mathrm{RR}}\!=\!0$). The dimensionless
sink energy is at $x_{D}=10^{5}$, approximately appropriate for the
tidal radius in the GC. Lines show $g(x,\tau)$ at intervals of $\Delta\tau\!=\!0.07$,
with the lowest curve indicating the initial DF (arbitrarily assumed
here to be $g(x,0)\propto x^{0.1}$). The system was integrated up
to $2T_{h}$; the steady state DF is reached after $\sim\!0.8T_{h}$.
For comparison, an $x^{1/4}$ power-law is also shown.}
\end{figure}

Here we explore the effect of RR on the steady state DF. We begin
by checking the numerical convergence of Eq. (\ref{e:dgdt}) to the
BW76 solution, $f(\mE)\!\propto\!\mE^{p}$ with $p\!=\!1/4$ for $\mE_{h}\!\ll\!\mE\!\ll\!\mE_{D}$,
which is obtained under the assumption that the $\mE$-dependent $J$-averaged
sink terms can be neglected, $R_{\mathrm{NR}}\!=\! R_{\mathrm{RR}}\!=\!0$,
and that the disruption of stars by the MBH can be expressed by the
boundary condition $f(\mE\!>\!\mE_{D})\!=\!0$. Figure (\ref{f:bw76})
confirms that the DF indeed converges to this solution on a relaxation
timescale, and we also verified that this holds irrespective of the
initial DF assumed. Rapid conversion was also confirmed by the $N$-body
simulations of Merritt \& Szell (\cite{MS05}). The BW76 DF corresponds
to a power-law density profile $n(r)\propto r^{-\alpha}$ with $\alpha\!=\! p+3/2\!=\!7/4$
(equation \ref{e:nr}). Since $T_{\textrm{NR}}(\mE)\propto P(\mE)/N(>\mE)$,
and $T_{\textrm{NR}}(\mE=\sigma_{h}^{2})=T_{h}$, the implied \emph{local}
NR time is

\begin{equation}
T_{\textrm{NR}}(\mE)=T_{h}\left(\mE/\sigma_{h}^{2}\right)^{3/2-\alpha}\,.\label{e:TrE}\end{equation}

\begin{figure}
\includegraphics[%
  width=1.0\columnwidth,
  keepaspectratio]{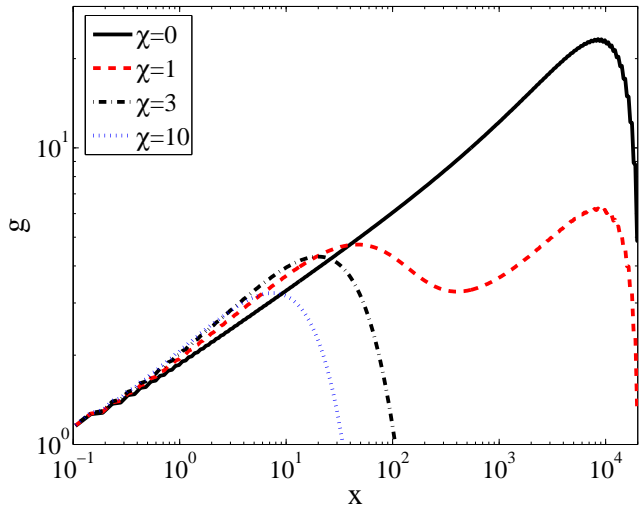}

\caption{\label{f:bwRR} Growth of a cusp in presence of RR, with the same
initial conditions as in Fig. (\ref{f:bw76}) in the presence of the
NR and RR sink terms for RR efficiency factors of $\chi\!=\!0,1,3,10$.
For RR efficiency factor $\chi\!\gtrsim\!3$, RR leads to an exponential
cutoff of the DF at high energies.}
\end{figure}

We now add back the NR and RR sink terms and solve the full problem.
Figure (\ref{f:bwRR}) shows the results of the steady state solution
of equation (\ref{e:dgdt}) for several values of $\chi$. The $\chi\!=\!0$
case (NR only) reproduces the solution studied by BW77, who showed
that the presence of the NR sink term does not change the power-law
behavior and index of the DF. We find, as expected, that for low energies
where RR is negligible, neither does the presence of the RR sink term.
At higher energies, the response of the DF depends on the efficiency
of RR. For $\chi\!\gtrsim\!10$, RR is so efficient that it exponentially
depletes the high energy end of the DF. However, for the likely efficiency
factor $\chi\!=\!1$, GR precession at high energies limits the efficiency
of RR (cf Fig. \ref{f:GC}), and the DF is not completely depleted,
but continues to rise toward $\mE\!\rightarrow\!\mE_{D}$ after a
moderate initial drop. 

\subsection{The effects of RR on the stellar current}

\label{ss:flow}

The solution of the Fokker-Planck equation determines the stellar
current $Q(x)$ into the MBH. This in turn determines the rates and
modes of stellar capture by the MBH. Since most of the contribution
to prompt infall events comes from orbits with energy $x\!\gtrsim\! x_{p}$,
and to GW inspiral from $x\!\gtrsim\! x_{\mathrm{GW}}$ (\S\ref{ss:losscone},
\S \ref{s:implications}), the current at these critical energies
determines the prompt infall and gravitational inspiral event rates.
In the GC, $x_{p}\!=\!2.7$ and $x_{GW}\!=\!58$ (\S\ref{ss:losscone}).

The rate at which stars flow toward the MBH is given by

\begin{equation}
I(\mE,t)=I_{0}Q(x,t),\label{e:I}\end{equation}
 where the dimensional current scale is set by

\begin{equation}
I_{0}\equiv{\frac{8\pi^{2}}{3\sqrt{2}}}r_{h}^{3}n_{h}{\frac{(G\Ms)^{2}\ln\Lambda n_{h}}{\sigma_{h}^{3}}}\sim{\mathcal{{O}}}\left(\frac{N_{h}}{T_{h}}\right)\,.\label{e:I0}\end{equation}
 For non-equilibrium DFs, $Q\!\sim\!1$, and there is a large current
$I\!\sim\! I_{0}$. The {}``sink-less'' BW76 solution (\S\ref{ss:DF})
has a steady state {}``zero-current'' ($0\!<\! Q\!\ll\!1$) solution
that is $\mE$-independent and is strongly suppressed by the bottle-neck
at $\mE_{D}$. The presence of the loss-cone, as expressed by the
NR and RR sink terms, modifies the DF at high energies and shifts
the effective high-energy boundary to lower energies. As a result,
$Q$ increases well above the zero-current solution and becomes $\mE$-dependent.

\begin{figure}
\includegraphics[%
  width=1.0\columnwidth,
  keepaspectratio]{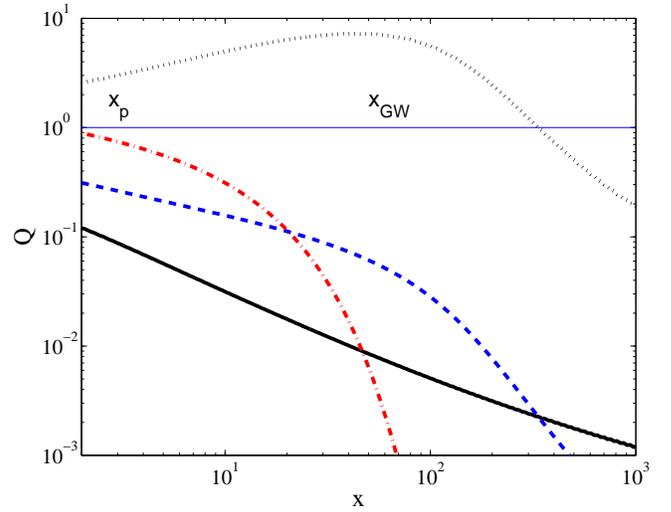}

\caption{\label{f:flow} The steady state current $Q$ as function of dimensionless
energy $x$, for various choices of the RR efficiency parameter $\chi$.
Solid line: $\chi\!=\!0$ (NR only). Dashed line: $\chi\!=\!1$. Dot-dashed
line: $\chi\!=\!10$. The capture rates are determined by the value
of the current at the critical energies (shown here at values typical
for the GC, \S\ref{sss:tiddisr}, \S\ref{sss:GW}). The dotted line
is the ratio $Q_{1}/Q_{0}$ between the currents of the $\chi\!=\!1$
and $\chi\!=\!0$ cases. At $x_{\mathrm{GW}}$, $Q_{1}/Q_{0}\!\sim\!8$,
so the GW inspiral rate is dominated by RR. At $x_{p}$, $Q_{1}/Q_{0}\!\sim\!3$
so prompt infalls are somewhat enhanced by RR, as found by RT96
and RI98. When $\chi\!=\!10$, RR is so efficient that it evacuates
a large fraction of the stars at energies higher than $x_{\mathrm{GW}}$,
and the GW inspiral rate is suppressed relative to the $\chi\!=\!1$,
whereas the prompt infall rate is further enhanced. }
\end{figure}

\begin{figure}
\includegraphics[%
  width=1.0\columnwidth,
  keepaspectratio]{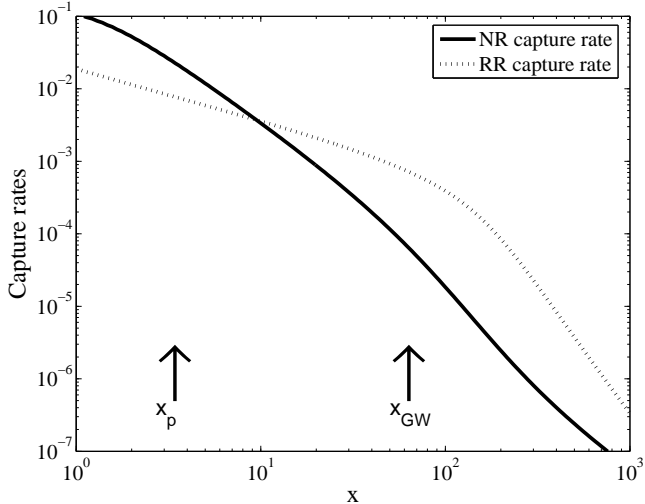}

\caption{\label{f:rates} The differential capture rates $x^{5/2}R_{\mathrm{NR}}$
(solid line) and $x^{5/2}R_{\mathrm{RR}}$ with $\chi\!=\!1$ (dashed
line) as function of dimensionless energy $x$. At high energies near
$x_{\mathrm{GW}}$, where GW sources originate, captures are dominated
by RR. Prompt infall is dominated by NR in the region near $x_{p}$,
in the transition between the full and empty loss-cone regimes.}
\end{figure}

Figure (\ref{f:flow}) shows the current $Q(x)$ for several values of
the RR efficiency factor $\chi$. The accelerated relaxation due to RR
increases the current above the values induced by NR only. For
$\chi\!=\!1$ the largest enhancement is attained at about the critical
energy for GW. When $\chi\!\gtrsim\!10$, RR is so efficient that the
current depletes the DF already at energies $x\!\lesssim\! x_{GW}$.
As a consequence, the current drops to zero at $x\!>\! x_{GW}$, and
therefore so does the GW inspiral event rate (which is $\propto
Q(x_{\rm GW})$, see eq. [\ref{e:GWJ}]). Figure (\ref{f:rates}) shows
separately the contributions of the NR and RR sink terms in
Eq. (\ref{e:dgdt}) to the total capture rate, and demonstrates that
far from the MBH (low-$\mE$), where most tidally disrupted stars
originate, NR dominates the capture rate, whereas at closer distances
(high-$\mE$), where GW inspiral stars originate, RR dominates the
capture rate (see also Fig. \ref{f:GC} below).

\section{Observable implications of resonant relaxation }

\label{s:implications}

Most previous estimates of the loss rates did not take into account
RR. Here we show that RR can strongly influence the GW inspiral rate,
and may play an important role in determining the dynamical structures
observed in the inner GC.

\subsection{Infall and inspiral processes}

\label{ss:inprocess}

\begin{figure}
\includegraphics[%
  width=1.0\columnwidth,
  keepaspectratio]{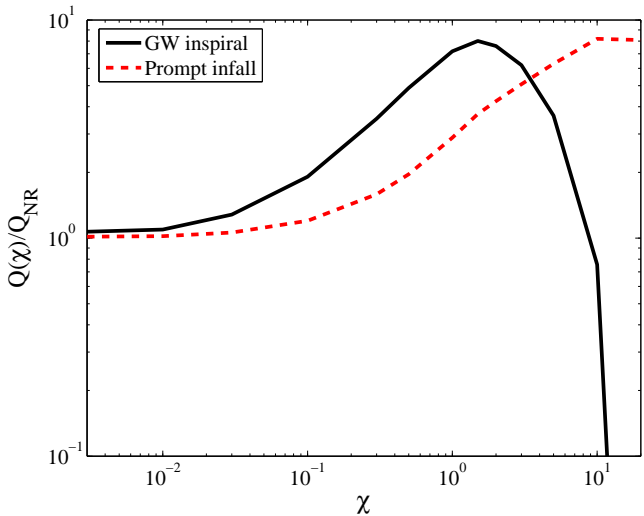}

\caption{\label{f:Chi} The $\chi$-dependence of the enhancement of the prompt infall and GW inspiral event rates by RR over NR, as expressed
by the stellar current at the critical energy $x_{c}$, $Q(x_{c};\chi)/Q(x_{c};0)$,
for $x_{c}\!=\! x_{p},x_{GW}$.}
\end{figure}

\subsubsection{Prompt infall}

\label{sss:tiddisr}

Figure (\ref{f:Chi}) shows the change in the estimated rates of prompt
infall and inspiral due to the inclusion of RR, as function of the
RR efficiency parameter $\chi$. RR does not significantly increases
the rate of prompt infall events. This result has already been
demonstrated by the $N$-body simulations of RT96 and RI98, and is
explained by the fact that most tidally disrupted stars originate
from low-$\mE$ orbits with $\mE\!\sim{\cal {O}}(\mE_{h})$ (Lightman
\& Shapiro \cite{LS77}; Baumgardt et al. \cite{Bea05}), where mass
precession suppresses scalar RR. Our analysis is complementary to
that of RT96 and RI98 in that they took into account relaxation in
$J$-space, but not in $\mE$-space, whereas ours explicitly solves
self-consistently the DF and star current in $\mE$-space, but does
not deal with the flow in $J$-space. Our results confirm the conclusion
that for the likely value of the RR efficiency factor $\chi\!=\!1$,
the tidal disruption rate is enhanced by RR by only a factor of $\sim\!3$
(Figs \ref{f:flow}, \ref{f:Chi}). We note, however, that for higher
values of $\chi$, the tidal disruption rate may increase significantly.

\subsubsection{Gravitational wave inspiral}

\label{sss:GW}

The detection of GW emission from compact remnants inspiraling into
MBHs (EMRIs), is one of the major goals of the planned space-borne
\textit{LISA} GW detector %
\footnote{http://lisa.jpl.nasa.gov/%
}. The predicted event rate per galaxy is very uncertain, as estimates
vary in the range ${\textrm{Gyr}}^{-1}\!\lesssim\!\Gamma_{\mathrm{GW}}\!\lesssim\!{\textrm{Myr}}^{-1}$
per galaxy (e.g. Hils \& Bender \cite{HB95}; Sigurdsson \& Rees \cite{SR97};
Ivanov \cite{IV02}; Freitag \cite{FR01}, \cite{FR03}; Hopman \&
Alexander \cite{HA05}).

Only stars originating from tightly bound orbits with energies $\mE\!\gtrsim\!\mE_{\mathrm{GW}}$
can complete their inspiral and reach a high-frequency orbit observable
by \emph{LISA}, without being scattered prematurely into the MBH or
to a wider orbit. An approximate upper-limit for the semi-major axis
from which inspiral is possible can be obtained by equating the inspiral
time $t_{\textrm{GW}}$ with $(1\!-\! e)T_{\textrm{NR}}$, the NR
timescale for scattering by $\Delta J\!\sim\! J$, (Hopman \& Alexander
\cite{HA05}),

\begin{equation}
a_{\mathrm{GW}}\!\equiv\!\left(\frac{8\sqrt{G\Mbh}E_{1}T_{h}}{\pi
 c^{2}}\right)^{2/3}\,,\quad
 E_{1}\!\equiv\!\frac{85\pi}{3\!\times\!2^{13}}\frac{\Ms
 c^{2}}{\Mbh}\,,\label{e:dgw}\end{equation} (the corresponding
 critical energy is $\mE_{\mathrm{GW}}\!=\!
 GM_{\bullet}/2a_{\textrm{GW}}$).  There are three phases in the
 orbital evolution of a star that ends up as a GW source. The first is
 a scattering-dominated phase in $J$-space, where energy losses are
 negligible. This is followed by a transition phase, when
 $J\!\sim\!\mathrm{few}\times\! J_{\mathrm{LSO}}$, GW dissipation
 becomes significant, and the inspiral time is comparable to the
 scattering time. Finally, the star spirals in by GW emission so
 rapidly that it effectively decouples from the gravitational
 perturbations of the background stars. The value of $a_{\textrm{GW}}$
 is determined in the second, transition phase by the interplay
 between angular momentum relaxation and GW dissipation. The
 $J$-values typical of the transition phase are also those where
 GR-precession quenches RR, at $J\!<\!
 J_{Q}\!\sim\!\mathrm{few\times\!}J_{\mathrm{LSO}}$
 (Eq. \ref{e:tGR}). As a result, $a_{\textrm{GW}}$ is still determined
 by NR through Eq. (\ref{e:dgw}) \emph{even when RR dominates the
 dynamics at} $J\!\gg\! J_{\mathrm{LSO}}$.

\begin{figure}
\includegraphics[%
  width=1.0\columnwidth,
  keepaspectratio]{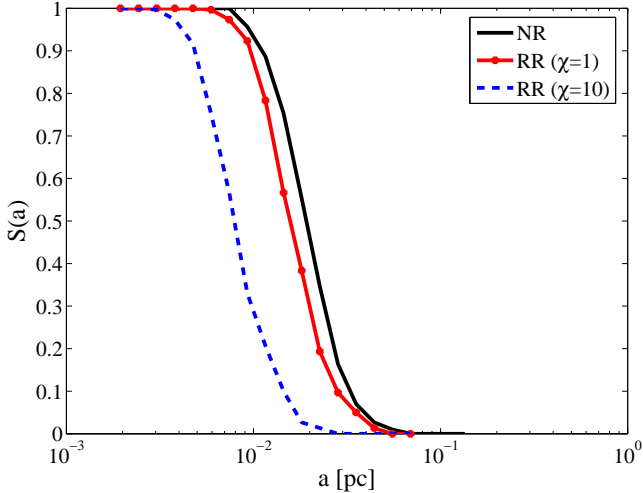}

\caption{\label{f:Sa} The fraction of stars $S(a)$ that produce a
detectable GW signal out of all stars consumed by the MBH, as function
of the initial semi-major axis. The solid line is for NR only; the
dots for NR and RR (including GR quenching near $J_{lc}$). The two
plots are almost identical, proving that RR does not affect the
inspiral probability. The dashed line is for NR and RR, but with an
artificially decreased GR precession (by a factor 10). In this case
$S(a)$ drops at slightly smaller values of $a$. The small difference
implies that $S(a)$ is not sensitive to the exact value of $J_{Q}$
where RR is quenched.}
\end{figure}

To verify that RR does not significantly affect the transition phase
of inspiral and the value of the critical energy, we followed the
approach used by Hils \& Bender (\cite{HB95}) and Hopman \& Alexander
(\cite{HA05}), who performed Monte Carlo (MC) simulations of the
scattering of stars in $J$-space. We compare the case where $T_{J}^{s}\!=\! T_{\mathrm{NR}}$
(as was assumed in Hils \& Bender \cite{HB95}; Hopman \& Alexander
\cite{HA05}) to the case where $T_{J}^{s}$ includes RR (Eq. \ref{e:TJ}).
Figure (\ref{f:Sa}) shows our results in terms of the inspiral fraction
$S(a)$, the fraction of MC experiments that end with the star spiraling
in, relative to the total, which also includes the experiments that
end with the star plunging directly into the MBH without emitting
a detectable GW signal (the third possibility of diffusion to lower
energies is not relevant here since only $J$-scattering is considered).
The critical semi-major axis corresponds to $S(a_{\mathrm{GW}})\!\sim\!0.5$.
Figure (\ref{f:Sa}) shows that RR has little effect on the value
of $a_{\mathrm{GW}}$, which is completely determined by NR. Furthermore,
the outcome is not sensitive to the exact value of $J_{Q}$ where
RR is quenched, as is demonstrated by the fact that $a_{\mathrm{GW}}$
is only slightly decreased when the strength of GR precession is artificially
lowered by a factor of 10 (by increasing $\chi$ to 10), which is
equivalent to assuming a smaller value of $J_{Q}$.

Because $S(a)$ is well approximated by a step function at $a_{\mathrm{GW}}$
(or equivalently, $S(\mE)$ is a step function at $\mE_{\mathrm{GW}}$),
the inspiral event rate per galaxy \emph{without RR} can be estimated
by (Hopman \& Alexander \cite{HA05})

\begin{equation}
\Gamma_{\textrm{NR}}^{\textrm{GW}}\simeq f_{s}\int_{\mE_{\mathrm{GW}}}^{\infty}\!\frac{\mathrm{d}\mE\, n(\mE)}{\ln(J_{c}/J_{\textrm{LSO}})T_{\textrm{NR}}(\mE)}\,,\label{e:SU}\end{equation}
 where $f_{s}$ is the population fraction of the specific GW sources
under consideration (e.g. white dwarfs). 

Figure (\ref{f:Chi}) shows that the GW inspiral event rate reaches its
maximal enhancement, of about an order of magnitude, for
$\chi\!\simeq\!1$ and is enhanced over a wide range of values. Only
when $\chi\!\gtrsim\!10$ is the RR depletion of high energy GW
inspiral candidates with $x\!\gtrsim\! x_{\mathrm{GW}}$ so strong
(Fig. \ref{f:bwRR}) that the GW inspiral rate falls below that
predicted for NR only. This depletion can be expressed by introducing
an effective high-energy cutoff $\mE_{\mathrm{RR}}\!\ll\!\mE_{D}$,
such that $n(\mE)\!\simeq\!0$ for $\mE\!>\!\mE_{\mathrm{RR}}$. Two
very different situations can occur, depending on whether the critical
energy $\mE_{\textrm{GW}}$ is larger or smaller than the cutoff energy
$\mE_{\mathrm{RR}}$.

If $\mE_{\textrm{GW}}\!\gg\!\mE_{\mathrm{RR}}$, the DF vanishes for
all energies where stars could complete their inspiral. In this case
the GW event rate is vanishingly small. Fortunately for the prospects
of EMRI detection, it appears likely that this is \textit{not} the
case (see Figs. \ref{f:flow}, \ref{f:rates}, \ref{f:Chi}). When
$\mE_{\textrm{GW}}\!\ll\!\mE_{R}$, the GW event rate is given by

\begin{equation}
\Gamma_{\mathrm{tot}}^{\textrm{GW}}\simeq\Gamma_{\textrm{NR}}^{\textrm{GW}}+f_{s}\int_{\mE_{\textrm{GW}}}^{\infty}\!\frac{\mathrm{d}\mE\, n(\mE)}{\bar{T}_{\textrm{RR}}^{s}(\mE)}=f_{s}I_{0}Q(x_{\textrm{GW}})\,,\label{e:GWJ}\end{equation}
 where the last equality follows from Eq. (\ref{e:dgdt}) in steady
state.

The Galactic MBH is the prototypical \emph{LISA} target (the orbital
frequencies around more massive MBHs are below the \emph{LISA} sensitivity
band). Applying our result to the GC, we estimate that RR increases
the predicted \textit{LISA} event rate per galaxy by up to a factor
(Fig. \ref{f:flow})

\begin{equation}
{\frac{\Gamma_{\mathrm{tot}}^{\textrm{GW}}}{\Gamma_{\textrm{NR}}}}={\frac{Q_{\textrm{RR}}(x_{\textrm{GW}})}{Q_{\textrm{NR}}(x_{\textrm{GW}})}}\simeq8\,,\label{e:gL}\end{equation}
 over previous predictions that assumed NR only. One caveat is that
 our analysis neglects the effects of mass segregation in a multi-mass
 population. For example, low-mass remnants will be driven by mass
 segregation to a flatter density distribution, which will decrease
 their contribution to the inspiral event rate (Hopman \& Alexander
 \cite{HA05}, \cite{HA06}), and the opposite will happen for high mass
 objects. Here we do not consider a multi-mass population.

\subsection{Resonant relaxation in the Galactic center}

\label{ss:GC}

At a distance of $\sim\!8$ kpc (Reid \& Bruthaler \cite{RB04}),
the Galactic MBH is the closest and most accessible MBH. Astrometric
and radial velocity measurements of stars closely orbiting the MBH
indicate that its mass is $\Mbh\!\sim\!(3\!-\!4)\!\times\!10^{6}\Mo$
(Sch\"{o}del et al. \cite{S02}; Ghez et al. \cite{Gh03}; Eisenhauer
et al. \cite{Ei05}; for a review see Alexander \cite{A05}). Observations
of the GC provide the most detailed information available about stars
close to a MBH, where the orbits are Keplerian. The Galactic MBH offers,
in principle, the best chances of confronting RR with observations.
We consider here the implications of RR for several observed and predicted
features of the GC: the young coherent star disks, the central randomized
cluster and the relaxed old giants; the orbital parameters of the
innermost stars; the stellar density distribution; and the hypothesized
dense stellar cluster of stellar black holes (SBHs).

\begin{figure}
\includegraphics[%
  width=1.0\columnwidth,
  keepaspectratio]{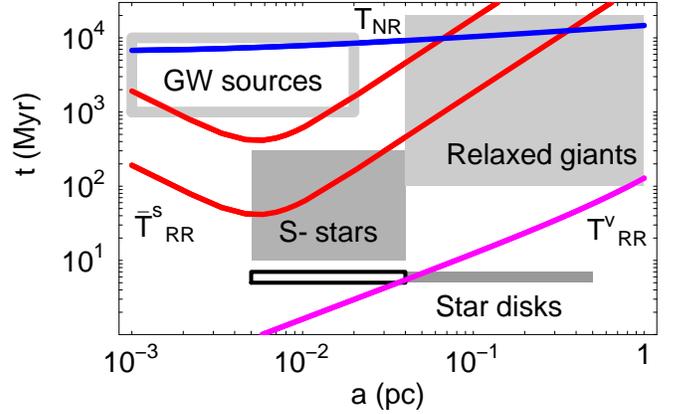}

\caption{\label{f:GC} Stellar components, timescales and distance scales
in the GC (equating $a\!\sim\! r$). The NR timescale $T_{\mathrm{NR}}$
(top straight line); the averaged scalar RR timescale $\bar{T}_{\textrm{RR}}^{s}$,
estimated for $1\,\Mo$ stars (top curved line) and $10\,\Mo$ stars
(bottom curved line); the vector RR timescale $T_{\mathrm{RR}}^{v}$
(bottom straight line); the position and estimated age of the young
stellar rings in the GC (filled rectangle in the bottom right); the
position and age of the S-stars if they were born with the disks (empty
rectangle in the bottom left); the position and maximal lifespan of
the S-stars (filled rectangle in the middle left); the position and
age of the dynamically relaxed old red giants (filled rectangle in
the top right); and the position and age of the reservoir of GW inspiral
sources, where the age is the progenitor's age or the time to sink
to the center (empty rectangle in the top left).}
\end{figure}

Figure (\ref{f:GC}) compares the distance scales and the ages or
lifespans of the various dynamical structures and components in the
inner pc of the GC with the relaxation timescales. The NR timescale
in the GC, which is roughly independent of radius, is $T_{\textrm{NR}}\!\sim\!\mathrm{few\!\times\!10^{9}}\,\mathrm{yr}$
(Eq. \ref{e:tr}). The scalar RR relaxation time $\bar{T}_{\mathrm{RR}}^{s}$
is shown for $\chi=1,10$, or, equivalently, for $\Ms=1,10\,\Mo$
with $\chi\!=\!1$. At large radii the RR time decreases towards the
center, but for small radii, where GR precession becomes significant,
it increases again (see \S\ref{s:RR}). The vector RR timescale $T_{\mathrm{RR}}^{v}$,
in contrast, decreases unquenched with decreasing radius. Structures
whose estimated age exceeds these relaxation timescales must be relaxed.
Structures whose lifespan exceeds the relaxation timescales may be
affected, unless we are observing them at an atypical time soon after
they were created. As is further discussed below, RR can naturally
explain some of the systematic differences between the various dynamical
components in the GC. 

The observed deviations from spherical symmetry of the potential in
the GC do not affect RR. As an example, one may consider a ring of
mass $M_r$ and radius $r$. Such a perturbation to the potential would
change orbits with semi-major axis $a$ on the Kozai time (Kozai
\cite{KO62}) $t_K\sim 2\pi(\Mbh/M_{r})(r/a)^3P(a)$, which is larger
than $t_M$ by a factor of order $t_{K}/t_{M}\sim2\pi
N(<a)\Ms/M_{r}(r/a)^3\leq2\pi N(<r)\Ms/M_{r}$. This implies that only
if the mass of the ring is comparable to the enclosed mass, such a
perturbation to sphericity will be important. This is not the case in
the GC. For example, the heaviest young stellar disk has a mass
$M_{r}<10^4\Mo$ (Nayakshin et al. \cite{Nay06}), giving $t_K/t_M>400$
for $r=0.4\pc$.

\subsubsection{RR and the young stars in the Galactic center}

\label{sss:youngstars}

Two distinct young stellar populations in the GC may be of particular
relevance for testing RR. At distances of $0.04$--$0.5$ pc from
the MBH there are about $\sim\!70$ young massive OB stars ($\Ms\!\gg\!10\,\Mo$,
lifespan of $t_{\star}\!=\!6\pm2$ Myr), which are distributed in
two nearly perpendicular, tangentially rotating disks (Levin \& Belobodorov
\cite{LB03}; Genzel et al. \cite{Gea03}; Paumard et al. \cite{P05}).
It appears that these stars were formed by the fragmentation of gas
disks (Levin \& Belobodorov \cite{LB03}; Levin \cite{L03}; Nayakshin
\& Cuadra \cite{NC04}; Nayakshin \& Sunyaev \cite{NS05}; Nayakshin
\cite{N06}). This young population co-exists with a relaxed population
of long-lived evolved giants ($t_{\star}\!>\!100\,\mathrm{Myr}$,
$\Ms\!\lesssim8\!\Mo$; Genzel, Hollenbach \& Townes \cite{Gen94}).
Inside the inner $0.04\,\mathrm{pc}$ the population changes. There
is no evidence for old stars, and the young stars there (the {}``S-stars'')
are main-sequence B-stars ($\Ms\!\lesssim15\,\Mo$, lifespans of $10^{7}\mathrm{\!\lesssim t_{\star}\!\lesssim\!2}\!\times\!10^{8}$
yr; Ghez et al. \cite{Gh03}; Eisenhauer et al. \cite{Ei05}) on randomly
oriented orbits with a random (thermal) $J$-distribution. The orbital
solutions obtained for a few of the S-stars indicate that they are
tightly bound to the MBH (Ghez et al. \cite{Gh03}; Eisenhauer et
al. \cite{Ei05}). There is to date no satisfactory explanation for
the presence of the S-stars so close to the MBH (see Alexander \cite{A05}
for a review). One possibility is that they originated far from the
MBH and were captured near it by dynamical exchange interactions (Gould
\& Quillen \cite{GQ03}; Alexander \& Livio \cite{AL04}). In that
case only their lifespan can be confidently determined, but not their
actual age. Another possibility is that they also originated in the
star disks. In that case their age is the same as that of the disks
and is much shorter than their lifespan. 

The existence of coherent dynamical structures in the GC constrains
the relaxation processes on these distance scales, since the
relaxation timescales must be longer than the structure age
$t_{\star}$ to avoid randomizing it. Figure (\ref{f:GC}) shows that
the observed systematic trends in the spatial distribution, age and
state of relaxation of the different stellar components of the GC are
consistent, and perhaps even caused by RR. The star disks are young
enough to retain their structure up to their inner edge at
$0.04\,\mathrm{pc}$, where $t_{\star}\!\sim\! T_{\mathrm{RR}}^{v}$ and
vector RR can randomize the disk. It is tempting to explain the
S-stars as originally being the inner part of the disk. However, RR
alone cannot explain why the S-stars are systematically less massive
than the disk stars. Regardless of their origin, their random orbits
are consistent with the effect of RR. Vector RR can also explain why
the evolved red giants beyond $~\!0.04\,\mathrm{pc}$, in particular
the more massive ones with
$t_{\star}\!\ll\!\min(T_{\mathrm{NR}},\bar{T}_{\mathrm{RR}}^{s})$ are
relaxed, since $T_{\mathrm{RR}}^{v}\!<\! t_{\star}$ out to $\sim\!1$
pc. Note that scalar RR is in itself not efficient enough to drain the
S-stars, but it can play a role in their non-zero eccentricities (see
also Levin \cite{L06}).

Rapid RR changes the orbital eccentricities. On the timescales
relevant for observations ($t\!\ll\! t_{\omega},$ Eq. \ref{e:tprec}),
the change grows linearly and could be detected, in principle, by
precision astrometry. However, the effect is small. In the linear
regime the shortest time-scale for changes in $J$ is determined by
scalar RR.  The fractional change per orbit is $\Delta
J_{P}/J_{c}\!\sim\!\sqrt{N(<\! a)}(\Ms/\Mbh)$ (Eq. \ref{e:Jw}), which
corresponds to a change in eccentricity of $\Delta
e_{P}\!\sim\!\sqrt{N(<\! a)}(\Ms/\Mbh)\sqrt{1-e^{2}}/e$.  For the best
measured orbit (the star S2 with $e\!=\!0.8760\!\pm\!0.0072$.
Eisenhauer et al. \cite{Ei05}) the predicted change per orbit is
$\Delta e_{P}\!\sim\!10^{-4}$ (assuming $\Ms=10\,\Mo$, $a=0.01\pc$ and
$N(<\! a)=2500$). This is beyond the current astrometric precision.

\subsubsection{RR and the stellar distribution in the Galactic center }

\label{sss:nGC}

\begin{figure}
\includegraphics[%
  width=1.0\columnwidth,
  keepaspectratio]{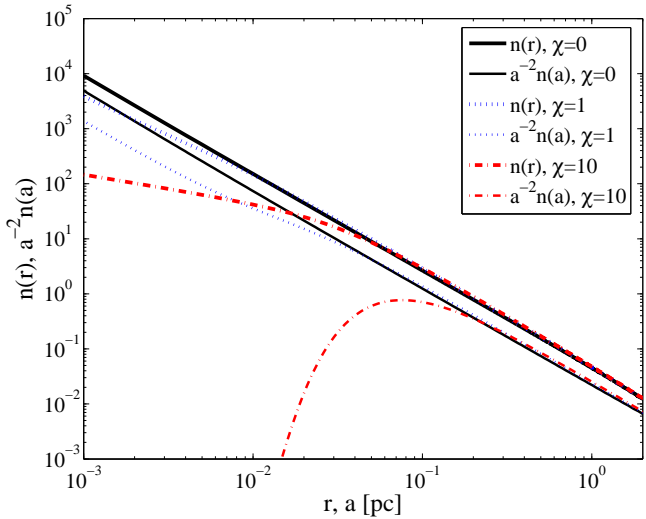}

\caption{\label{f:nGC} The density profile $n(r)$ and the distribution of
the semi-major axis $n(a)$ in the GC for several values of $\chi$.}
\end{figure}

Figure (\ref{f:nGC}) shows the density distribution $n(r)$ (Eqs.
\ref{e:nr}, \ref{e:nru}) and the distribution of semi-major axis
$a^{-2}n(a)$ (Eq. \ref{e:na}) for a model of the GC with different
values of $\chi$. The central depletion in the distribution of the
semi-major axis (or equivalently, energy) does not appear in the space
density profile, because unbound and weakly bound stars on eccentric
orbits spend a fraction of their orbit in the center. The effect of
RR on the density distribution only appears as a gradual flattening
of the cusp slope to $n_{u}(r)\!\propto\! r^{-1/2}$ (Eq. \ref{e:nru}).
The effect on the observed projected surface density distribution
will be even less noticeable. RR is predicted to flatten the density
distribution only at $\lesssim\!\mathrm{few\!\times\!10^{-3}\, pc}$,
for $\chi\!=\!1$. Farther out the distribution is virtually identical
with the NR solution. Since there are no observational data on the
density distribution for distances $\lesssim\!5\!\times\!10^{-3}\pc$
(Sch\"{o}del et al. \cite{S05}), it is presently not possible to
test this aspect of RR in the GC. However, if $\chi\!\gg\!1$, the
space density will flatten farther out, on a length-scale that may
be observationally accessible. It should be emphasized that any detailed
comparisons between the observed stellar distribution and models will
likely be difficult (Alexander \cite{A99}, \cite{A05}). The observed
density profile in the inner $0.4$ pc falls as $n(r)\!\propto\! r^{-1.4}$
(Genzel et al. \cite{Gea03}), less steeply than predicted by the
single mass BW76 solution. It is highly likely that additional dynamical
processes beyond the simple picture considered here are at work, such
as mass segregation (BW77). Furthermore, a large fraction of this
observed distribution is comprised of massive stars with lifespans
shorter than either the NR or RR relaxation times (Fig. \ref{f:GC});
these stars probably do not trace the relaxed population.

We note that if the power-law profile were to continue to very small
radii, the smallest radius $r_{1}$ where a star is still statistically
expected to be found, $r_{1}\!=\!
N_{h}^{1/(3-\alpha)}r_{h}\!\sim\!6\!\times\!10^{-5}\pc$ (for
$\alpha\!=\!1.4$), would be well inside radius where RR depletes the
cusp. If such a yet undetected population of tightly bound stars
exists, it should exhibit an exponential central suppression at
$\sim\!10^{-3}\,\mathrm{pc}$.

\subsubsection{RR and dark mass in the Galactic Center}

\label{sss:DM}

It is likely that the Galactic MBH is surrounded by a dynamically
significant distribution of dark mass in the form of compact remnants,
in particular SBHs that have sunk to the center by mass-segregation
over the lifetime of the Galaxy (Morris \cite{M93}; Miralda-Escud\'e
\& Gould \cite{MG00}; Hopman \& Alexander \cite{HA06}; Freitag,
Amaro-Seoane \& Kalogera \cite{Fre06}). The existence of dark matter
can be constrained by detecting orbital deviations from purely
Keplerian motions (Mouawad et al. \cite{MO04}; Alexander
\cite{A05}). The co-existence of a dense cluster of SBHs with the
dense stellar cusp may have interesting dynamical implications
(e.g. Alexander \& Livio \cite{AL04}). Efficient RR could deplete this
dark mass component by rapidly draining the compact remnants into the
MBH. For our assumed RR efficiency factor $\chi\!=\!1$ the depletion
will be significant only on scales
$<\!\mathrm{few\!\times\!10^{-3}}\,\mathrm{pc}$ (Fig. \ref{f:nGC})
with negligible dynamical implications for the observed
stars. However, if $\chi\gtrsim10$, the inner $0.01\pc$ will be
depleted. Future high resolution observations of stellar orbits may
therefore further constrain RR.

RR may also affect the emission of gamma rays by annihilation of hypothetical
dark matter (DM) particles near MBHs. It was shown that gravitational
scattering between DM and stars leads to a DM density profile $\rho_{\textrm{DM}}\propto r^{-3/2}$
(Gnedin \& Primack \cite{GP04}; Merritt \cite{M04}). This profile
implies an annihilation rate which is proportional to $\log(r_{h}/r_{\textrm{in}})$.
In presence of RR, the inner cut-off of the cusp $r_{\textrm{in}}$
will be large as compared to the situation discussed in the literature
where RR is absent, thus somewhat decreasing the annihilation rate.

\section{Discussion and summary }

\label{s:summary}

RR is a coherent relaxation process that operates when symmetries
in the gravitational potential restrict the evolution of the orbits
(R96; RI98; Tremaine \cite{T05}). As a result, the orbits maintain
their orientation over many orbital periods, and over that time the
stars exert coherent mutual torques. These efficiently change the
orbital angular momentum $\mathbf{J}$ (energy relaxation continues
on the slow NR timescale). RR operates only under specific conditions\@.
However, when these are met, it can be orders of magnitude more efficient
than NR. Here we consider \emph{scalar RR}, which affects both the
direction and magnitude of $J$, in the regime around a MBH where
the orbits are Keplerian (where both mass and GR precession are negligible),
and \emph{vector RR}, which affects the direction but not the magnitude
of $J$, in the regime around a MBH where the potential is spherical.
Scalar RR can deflect stars into the MBH, whereas vector RR can only
change their orbital orientation.

In this paper we studied the effect RR has on the energy DF of stars
near a MBH. We explored the consequences for the disruption and capture
of stars by the MBH, in particular through inspiral by GW emission,
and used our results to interpret the properties of the observed dynamical
components in the GC. The complex full 2+1 $(\mE,J;t)$ problem was
reduced to an approximate 1+1 $(\mE;t)$ Fokker-Planck equation, which
we solved numerically. We also carried out Monte Carlo experiments
to test some of our assumptions.

Our results are as follows. (1) RR leads to the depletion of the high-energy
end of the DF (stars tightly bound to the MBH), accompanied by an
enhanced current of stars to high energies. The exact extent of the
depletion and the effective high-energy cutoff depend on the poorly
determined efficiency of RR. The currently available estimates of
the RR efficiency factor indicate that $\chi\!\sim\!1$. (2) We confirm
the result of RT96 that for $\chi\!=\!1$, the direct tidal disruption
rate is only modestly enhanced by RR. This is because scalar RR is
most efficient close to the MBH where the orbits are Keplerian, whereas
most of the tidally disrupted stars originate farther away from the
MBH. A higher RR efficiency than assumed here could substantially
\emph{increase} the tidal disruption rate. (3) We show, in contrast,
that the GW inspiral rate is dominated by RR dynamics, and is likely
increased by almost an order of magnitude relative to the rate predicted
assuming NR only. This is because stars undergoing GW inspiral originate
very close to the MBH. A higher RR efficiency than assumed here could
substantially \emph{decrease} the GW inspiral rate by completely depleting
the tightly bound stars. (4) We apply our results to the GC and show
that vector RR can naturally explain the inner cutoff of the ordered
star disks at $0.04$ pc and the transition to the randomized inner
S-star cluster, as well as the randomized state of the old red giants
in the inner $\sim\!1$ pc. We also show that scalar RR is consistent
with the presence of the disks, as it is too slow to disrupt them.
If the S-stars were born in the disk on circular orbits, then scalar
RR may explain may explain their present non-zero eccentricities.

We note that there are additional processes that could be affected
by RR, which we did not study here. Stars that undergo tidal capture
and subsequent tidal heating ({}``squeezars'', Alexander \& Morris
\cite{AM03}) also originate close to the MBH, where RR dominates
the dynamics. Squeezars could be directly observed in the GC if RR
substantially enhances their rates. A similar process is the tidal
capture of a binary companion by an intermediate mass black hole in
a young cluster and the subsequent Roche-lobe feeding that could power
an ultra-luminous X-ray source (Hopman, Portegies Zwart \& Alexander
\cite{HPZA04}; Hopman \& Portegies Zwart \cite{HPZ05}; Baumgardt
et al. \cite{Bea05}). The role of RR in tidal capture is still unclear,
as there is no RR quenching by GR precession at the tidal capture
radius that can stop the star from being rapidly destroyed. 

There are several limitations and uncertainties in our results that
will have to be addressed by future studies. Our treatment of the
problem in $\mE$-space incorporates the RR sink terms in approximate
form only. The efficiency of RR is poorly determined as it has been
calibrated based on a restricted set of $N$-body simulations and its
dependence on the parameters of the stellar system has not been
investigated in full. This uncertainty is significant, as it could
potentially reverse the sign of the effect of RR on the GW inspiral
rate. While the rate is increased by an order of magnitude if
$\chi\!=\!1$, it is completely suppressed if $\chi\!\gtrsim\!10$
(Fig. \ref{f:Chi}).  An important omission in our treatment is the
assumption of a single mass stellar population. A multi-mass
population will induce mass segregation. This will modify the DF
directly (BW77), and likely also change the RR efficiency. The stellar
DF also depends on processes that destroy one type of star while not
affecting others, such as stellar collisions (e.g. Freitag \& Benz
\cite{FB01}; \cite{FB05}).

Progress on these issues will likely require large scale numerical
simulations. The effects of RR on the DF of stellar systems with MBHs
have not yet been studied by $N$-body simulations. The recent fully
self-consistent $N$-body simulations of Baumgardt et al. (\cite{Baum04a},
\cite{Baum04b}), Preto et al. (\cite{P04}) and Merritt \& Szell
(\cite{MS05}), indicate that such studies are almost within reach
of current hardware. Such simulations are particularly important for
predicting GW inspiral rates, as these depend on a combination of
complex mechanisms, including mass-segregation and RR, which are hard
to assess with (semi-) analytical methods (e.g. Baumgardt et al. \cite{Bea05}).
Our analysis shows that it may be of considerable importance to use
a relativistic potential in such simulations (see e.g. Rauch \cite{R99}),
since otherwise it is unclear whether RR is quenched, and a large
number of stars may fall directly into the MBH without emitting an
observable GW signal. Yet larger-scale modeling will require abandoning
the exact $N$-body approach in favor of approximate methods such
as Monte Carlo simulations with RR (Freitag \& Benz \cite{FB01},
\cite{FB02}) or numerical 2+1 Fokker-Planck models.

We note that throughout we assumed a spherically symmetric
DF. Deviations from spherical symmetry may affect both RR and the
loss-cone structure in phase-space (see e.g. Magorrian \& Tremaine
[\cite{MT99}], who discuss the possibility of a loss 'wedge' in case
of a triaxial DF). Typically this enhances the event rates.

Finally, we note that the observed dynamics of the GC provide a
promising empirical basis for calibrating and cross-checking
theoretical and numerical studies of RR.


\acknowledgements{We thank K. Rauch for comments on the numerical results of RT96 and
A. Gualandris for her adaption of the $N$-body code by J. Makino
and P. Hut (http://www.artcompsci.org). TA is supported by ISF grant
295/02-1, Minerva grant 8484, and a New Faculty grant by Sir H. Djangoly,
CBE, of London, UK.}

\appendix{}

\section{\thesection{}. The logarithmic form of the Fokker Planck equation}

\label{s:LogE}

Because of the large ratio of the tidal radius and the radius of influence,
the natural way to integrate the Fokker-Planck equation is to divide
the energy range into equal logarithmic intervals. For convenience
we give here the equations in terms of the logarithmic distance variable
$z=\ln(1+x)$. The Fokker-Planck equation (\ref{e:dgdt}) without
sink terms is then written as

\begin{equation}
{\frac{\partial g(z,\tau)}{\partial\tau}}=-(e^{z}-1)^{5/2}e^{-z}{\frac{\partial Q(z,\tau)}{\partial z}},\end{equation}
 where $Q\equiv\Sigma_{i=1}^{4}Q_{i}(z,\tau)$ with\begin{eqnarray}
Q_{1}(z,\tau) & = & g(z,\tau)^{2}(e^{z}-1)^{-3/2}\,,\nonumber \\
Q_{2}(z,\tau) & = & g(z)\int_{z}^{z_{D}}dw(e^{w}-1)^{-3/2}{\frac{\partial g(w,\tau)}{\partial w}}\,,\nonumber \\
Q_{3}(z,\tau) & = & -e^{-z}(e^{z}-1)^{-3/2}{\frac{\partial g(z,\tau)}{\partial z}}\int_{-\infty}^{z}dwe^{w}g(w)\,,\nonumber \\
Q_{4}(z,\tau) & = & -e^{-z}{\frac{\partial g(z,\tau)}{\partial z}}\int_{z}^{z_{D}}dw(e^{w}-1)^{-3/2}e^{w}g(w)\,.\end{eqnarray}
 The logarithmic expressions for the sink terms can be included directly
from Eq. (\ref{e:dgdt}) by replacing $x\rightarrow e^{z}-1$. Written
this way, the numerical integration is simple and can be easily done
over many orders of magnitude in energy.

\section{\thesection{}. The non-resonant loss-cone sink term}

For convenience we reproduce here Eq. (14) of BW77, which interpolates
between the empty and the full loss-cone regimes of the NR sink term,

\begin{equation}
R_{\textrm{NR}}(x,\tau)={\frac{0.5g(x,\tau)x^{5/2}}{\alpha\ln(\Mbh/\Ms)\left[5.56+\ln(x_{D}/4x)/q(x,\tau)\right]}}\,,\label{e:NRBW}\end{equation}
 where

\begin{equation}
\alpha\equiv\left(\Ms/\Mbh\right)^{2}n_{h}r_{h}^{3}x_{D}\,,\qquad q(x,\tau)\equiv1.6\ln\left(6\Mbh/\pi\Ms\right)\alpha x^{-5/2}g(x,\tau)\,.\end{equation}


\begin{thebibliography}{ 2004b}
\bibitem[ 1999]{A99}Alexander, T., 1999, \apj, 520, 137 
\bibitem[ 2003]{AH03}Alexander, T., \& Hopman, C., 2003, \apj, 590, L29 
\bibitem[ 2003]{AM03}Alexander, T., \& Morris, M., 2003, \apj, 590, L25 
\bibitem[ 2004]{AL04}Alexander, T., \& Livio, M., 2005, \apj, 606, L21 
\bibitem[ 2005]{A05}Alexander, T., 2005, Physics Reports, 419(2--3), 65 
\bibitem[ 1976]{BW76}Bahcall, J. N., \& Wolf, R. A., 1976, \apj, 209, 214 (BW76) 
\bibitem[ 1977]{BW77}Bahcall, J. N., \& Wolf, R. A., 1977, \apj, 216, 883 (BW77)
\bibitem[ 2004]{BC04}Barack, L., \& Cutler, C., 2004, PRD, D70, 122002 
\bibitem[ 2004a]{Baum04a}Baumgardt, H., Makino, J., \& Ebisuzaki, T., 2004a, \apj, 613, 1133 
\bibitem[ 2004b]{Baum04b}Baumgardt, H., Makino, J., \& Ebisuzaki, T., 2004b,\apj, 613, 1143 
\bibitem[ 2005]{Bea05}Baumgardt, H., Hopman, C., Portegies Zwart, S. F., \& Makino, J.,
2005 (astro-ph/0511752) 
\bibitem[ 1987]{BT87}Binney, J. \& Tremaine, S., 1987, Galactic Dynamics (Princeton: Princeton
Univ. Press) 
\bibitem[ 1943]{Ch43}Chandrasekhar, S., 1943, Reviews of Modern Physics, 15, 1 
\bibitem[ 1978]{CK78}Cohn, H., \& Kulsrud, R. M. 1978, \apj, 226, 1087 
\bibitem[ 2005]{Ei05}Eisenhauer, et al., 2005, ApJ, 628, 246 
\bibitem[ 2000]{FM00}Ferrarese, L., \& Merritt, D., 2000, \apj, 539, L9 
\bibitem[ 1976]{FR76}Frank, J., Rees, M. J., 1976, MNRAS, 176, 633 
\bibitem[ 2001]{FB01}Freitag, M., \& Benz, W, 2001, A\&A, 375, 711 
\bibitem[ 2002]{FB02}Freitag, M. \& Benz, W., 2002, A\&A, 394, 345 
\bibitem[ 2001]{FR01}Freitag, M., 2001, Class. Quantum Grav., 18, 4033 
\bibitem[ 2003]{FR03}Freitag, M., 2003, \apj, 583, L21 
\bibitem[ 2002]{FB05}Freitag, M. \& Benz, W., 2005, MNRAS, 358, 1133
\bibitem[ 2006]{Fre06}Freitag, M., Amaro-Seoane, P., \& Kalogera, V., pre-print: astro-ph/0603280 
\bibitem[ 2004]{Gea04}Gair, J. R., Barack, L., Creighton, T., Cutler, C., Larson, S., L.,
Phinney, E. S., Vallisneri, M., 2004, Class. Quant. Grav. 21, S1595 
\bibitem[ 2000]{Geb00}Gebhardt, K., et al., 2000, \apj, 539, L13 
\bibitem[ 2003]{Gea03}Genzel, R. et al., 2003, \apj, 594, 812 
\bibitem[1994]{Gen94}Genzel, R., Hollenbach, D. \& Townes, C. H., 1994, Rep. Prog. Phys,
57, 417
\bibitem[ 2003]{Gh03}Ghez, A. M., et al., 2003, \apj, 586, L127 
\bibitem[ 2004]{GP04}Gnedin, O. Y., \& Primack, J. R., 2004, PRL, 93, 1302 
\bibitem[ 2003]{GQ03}Gould, A., \& Quillen, A., 2003, ApJ, 592, 935 
\bibitem[ 1995]{HB95}Hils, D., \& Bender, P. L., 1995, \apj, 445, L7 
\bibitem[ 2004]{HPZA04}Hopman, C., Portegies Zwart, S.F., \& Alexander, T., 2004, \apj,
604, L101 
\bibitem[ 2005]{HA05}Hopman, C., \& Alexander, T., 2005, ApJ, 629, 362 
\bibitem[ 2006]{HA06}Hopman, C., \& Alexander, T., 2006, pre-print: astro-ph/0603324
\bibitem[ 2005]{HPZ05}Hopman, C., \& Portegies Zwart, S. F., 2005, MNRAS, 363, L56 
\bibitem[ 2002]{IV02}Ivanov, P. B., 2002, \mnras, 336, 373, 2002 
\bibitem[ 1962]{KO62}Kozai, Y., 1962, \aj, 67, 591 
\bibitem[ 2003]{LB03}Levin, Y., \& Belobodorov, A. M., 2003, \apj, 596, 314 
\bibitem[ 2003]{L03}Levin, Y., 2003, (astro-ph/0307084) 
\bibitem[ 2006]{L06}Levin, Y., 2006, MNRAS, submitted
\bibitem[ 1977]{LS77}Lightman, A. P., Shapiro, S. L., 1977, ApJ, 211, 244 
\bibitem[ 1999]{MT99}Magorrian, J., Tremaine, S., 1999, MNRAS,309, 447 
\bibitem[ 1979]{MS79}Marchant, A.B., \& Shapiro, S.L., 1979, ApJ, 234, 317 
\bibitem[ 1980]{MS80}Marchant, A.B., \& Shapiro, S.L., 1980, ApJ, 239, 685 
\bibitem[ 2004]{M04}Merritt, D., 2004, PRL, 92, 1304 
\bibitem[ 2004]{MP04}Merritt, D., \& Poon, M. Y., 2004, ApJ, 606, 788 
\bibitem[ 2005]{MS05}Merritt, D., \& Szell, A., 2005 (astro-ph/0510498) 
\bibitem[ 2005]{Mea05}Miller, M. C., Freitag, M., Hamilton, D. P., \& Lauburg, V. M., 2005,
ApJ, 631, L117 
\bibitem[ 2000]{MG00}Miralda-Escud\'{e}, J., \& Gould, A., 2000, \apj, 545, 847 
\bibitem[ 1993]{M93}Morris, M., 1993, ApJ, 408, 496 
\bibitem[ 2004]{MO04}Mouawad, N., Eckart, A., Pfalzner, S., Sch\"{o}del, R., Moultaka,
J., Spurzem, R., 2004, Astronomische Nachrichten, Vol. 326, 2, 83-95 
\bibitem[ 1991]{MCD91}Murphy, B. W., Cohn, H. N., \& Durisen, R. H., 1991, ApJ, 370, 60 
\bibitem[ 2004]{NC04}Nayakshin, S., \& Cuadra, J., 2004, A\&A, 437, 437 
\bibitem[ 2005]{NS05}Nayakshin, S., Sunyaev, 2005 (astro-ph/0507687) 
\bibitem[ 2006]{Nay06}Nayakshin, S., Dehnen, W.,  Cuadra, J., \& Genzel, R., 2006, MNRAS, 366, 1410
\bibitem[ 2006]{N06}Nayakshin, S., 2006 (astro-ph/0512255) 
\bibitem[ 2005]{P05}Paumard, T., et al., 2005, \apj, submitted 
\bibitem[ 1964]{Pe64}Peters, P. C. 1964, Phys. Rev., 136, 1224 
\bibitem[ 2004]{P04}Preto, M., Merritt, D., Spurzem, R., 2004, \apj, 613, L109 
\bibitem[ 1996]{RT96}Rauch, K. P., \& Tremaine, S., 1996, New Astronomy, 149 (RT96) 
\bibitem[ 1999]{R99}Rauch, K. P., 1999, ApJ, 514, 725 
\bibitem[ 1998]{RI98}Rauch, K. P., \& Ingalls, 1998, MNRAS, 299, 1231 (RI98) 
\bibitem[ 2004]{RB04}Reid, M. J., \& Brunthaler, A., \apj, 616, 872 
\bibitem[ 2002]{S02}Sch\"{o}del, R. et al., 2002, \nat, 419, 694 
\bibitem[ 2005]{S05}Sch\"{o}del, R. et al., 2005, in prep. 
\bibitem[ 1979]{SM79}Shapiro, S.L., \& Marchant, A.B., 1979, ApJ, 225, 603 
\bibitem[ 1997]{SR97}Sigurdsson, S., and Rees, M. J., 1997, \mnras 284, 318 
\bibitem[ 1999]{SU99}Syer, D., Ulmer, A., 1999, MNRAS, 306, 35 
\bibitem[ 2002]{Tr02}Tremaine, S., et al., 2002, \apj, 574, 740 
\bibitem[ 2005]{T05}Tremaine, S., ApJ, 625, 143 
\bibitem[ 2004]{WM04}Wang, J., Merritt, D., 2004, ApJ, 600, 149 
\end{thebibliography}
\end{document}